\documentclass[preprint]{revtex4}  % Hochlademodus
\usepackage{amsfonts}
\usepackage{amsmath}
\usepackage{cancel}
\usepackage{graphicx}
\usepackage{mathcomp}
\usepackage[dvips]{epsfig}
\usepackage{color}
\usepackage{psfrag}

\newcommand{\ds}{\displaystyle}

\newcommand{\N}{\mathbb{N}}

\newcommand{\p}{\partial}
\newcommand{\n}{\nabla}
\newcommand{\di}{\nabla\cdot}
\newcommand{\be}{\begin{enumerate}}
\newcommand{\ee}{\end{enumerate}}
\newcommand{\bt}{\begin{tabular}}
\newcommand{\et}{\end{tabular}}

\newcommand{\ra}{\rightarrow}

\newcommand{\epsi}{\varepsilon}
\newcommand{\vp}{\varphi}
\newcommand{\vth}{\vartheta}

\newcommand{\qkq}{\quad,\quad}
\newcommand{\bra}[1]{\left\langle{#1}\right|}
\newcommand{\ket}[1]{\left|{#1}\right\rangle}
\newcommand{\braket}[2]{\left\langle{#1} \,|\, {#2} \right\rangle}
\newcommand{\norm}[1]{\left|\left| #1 \right|\right|}
\newcommand{\op}[1]{\mathsf #1}
\newcommand{\set}[1]{\mathcal{#1}}

\newcommand{\vecthree}[3]{\begin{pmatrix} #1 \\[1ex] #2\\[1ex] #3 \end{pmatrix}}

\renewcommand{\vec}[1]{\mbox{\boldmath $ #1 $}}

\begin{document}
\title{Active plasma resonance spectroscopy: \\
Eigenfunction solutions in spherical geometry}
\author{J. Oberrath}
%\author{T. Mussenbrock}
\author{R.\,P. Brinkmann}
\affiliation{Institute for Theoretical Electrical Engineering, 
Ruhr University Bochum,
D-44780 Bochum, Germany}
\date{\today}

\begin{abstract}
The term \textit{Active Plasma Resonance Spectroscopy} (APRS) denotes a class of related techniques which utilize, for~diagnostic purposes, 
the natural ability of plasmas to resonate on or near the electron plasma frequency $\omega_{\rm pe}$: A radio frequent signal (in the GHz range) is coupled into the plasma via an antenna or probe, the spectral response is recorded, and a mathematical model 
is used to determine plasma parameters like the electron density or the electron temperature. Based on the cold plasma model, this manuscript provides the general analytic expression of the electrical admittance of a spherical shaped probe immersed into a plasma. It is derived from the matrix representation of an appropriate operator, which describes the dynamical behavior of the probe-plasma system. This dynamical operator can be split into a conservative operator and a dissipative operator. It can be shown that the eigenvalues of the conservative operator represent the resonance frequencies of the probe-plasma system which are simply connected to the electron density. As an example, the result is applied to the spherical impedance probe and the multipole resonance probe. 
\end{abstract}

\maketitle

\section{Introduction}
The term ``active plasma resonance spectroscopy'' denotes a number of
similar plasma diagnostic methods, which exploit the natural ability 
of plasmas to resonate at or close to the plasma frequency of electrons.
The principle is quite simple: An electrical signal in the GHz range
is coupled to the plasma via an electrical probe. The spectrum of the
response of the plasma is recorded, and then evaluated based on a
specific mathematical model. From the structure of the spectrum one
is able to calculate the electron density and maybe other plasma
parameters. Clearly, the quality of the method strongly depends
on the quality and reliability of the model in use. 

One particular class of the method is that of electrostatic probes
\cite{takayama1960, messiaen1966, waletzko1967, vernet1975,
blackwell2005, sugai1999, scharwitz2009, lapke2008}. Within this class the coupling of a surface wave to the plasma is utilized, which excites resonance modes of frequencies below the electron plasma frequency. A number of different
approaches has been reported to provide a deeper understanding of the
resonance behavior of the system \cite{fejer1964, harp1964, 
crawford1964, dote1965, kostelnicek1968, cohen1971, tarstrup1972,
aso1973, bantin1974, booth2005, walker2006, lapke2007, xu2009,
xu2010, li2010, liang2011}. All of them are based on a fluid dynamical approach
and restricted to specific probe designs. 

However, a more general description has been provided by members of our own group \cite{lapke2013}. There, the whole class of electrostatic probes is analyzed by means of functional analytic (Hilbert space) methods. These methods are particular suited because explicit informations on the probe geometry or the electrical operation do not enter in the description. The most important result is given by the interpretation of a complex term as the electrical admittance of the probe-plasma system. The complex term refers to the resolvent of the dynamical operator that describes the system. This interpretation relies on a justified comparison to the admittance of an electrical lumped element series resonator. 

In this paper, the general expression of the electrical admittance of a spherical shaped probe is derived based on the matrix representation of the dynamical operator. This operator can be split into a conservative and a dissipative operator. We show that the eigenvalues of the conservative operator represent the resonance frequencies of the probe-plasma system. These resonance frequencies are simply connected to the electron plasma density. As an example, the result is applied to a) the spherical impedance probe (IP) investigated by Blackwell et al. \cite{blackwell2005} and b) the multipole resonance probe (MRP) \cite{lapke2008}.

\pagebreak 

\section{Model of an electrostatic probe of arbitrary shape}

\begin{figure}[h!]
\psfrag{electrodes (E)}{electrodes ($\set E_n$)}
\psfrag{sheath (S)}{sheath ($\set S$)}
\psfrag{isolator (I)}{isolator ($\set I$)}
\psfrag{dielectric (D)}{\hspace{-4mm} dielectric ($\set D$)}
\psfrag{plasma (P)}{plasma ($\set P$)}
\psfrag{holder}{holder}
\psfrag{vacuum seal}{vacuum seal}
\psfrag{shielded cable}{shielded cable}
\psfrag{chamber wall}{chamber wall}
\psfrag{signal generation/}{signal generation/}
\psfrag{evaluation}{evaluation}
\includegraphics[width= 0.8\columnwidth]{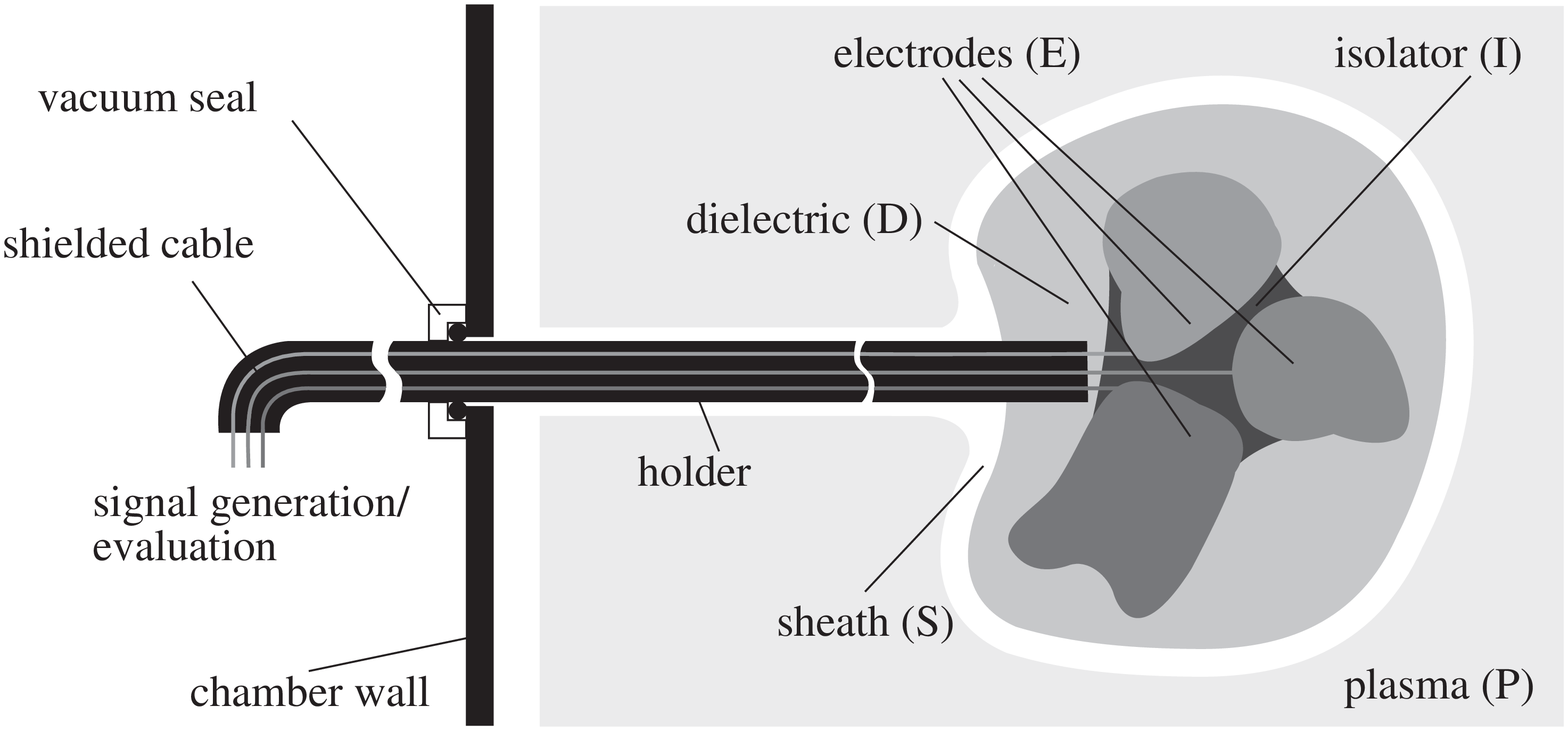}
\caption{Illustration of the abstract model for a N-electrode system \cite{lapke2013} (reprint). The electrodes are shielded to each other and the plasma by some dielectric medium. The whole probe, surrounded by a plasma sheath is immersed in a plasma volume.}\label{abstractmodel}
\end{figure}

In a recent paper a general model of an electrostatic probe is derived 
and analyzed \cite{lapke2013}. For details we refer this work. Here,
we summarize the most important aspects and results: The plasma chamber
(see figure \ref{abstractmodel}) is given as a simply connected, spatially 
bounded domain $\set V$, most of which is plasma (a simply connected 
subdomain $\set P$). Other subdomains of $\set V$ are the plasma boundary 
sheath $\set S$, which shields the plasma from all material objects, 
and possibly dielectric domains $\set D$. The boundary $\partial \set V$ 
of the domain $\set V$ is either grounded ($\set G$) or ideally 
insulating ($\set I$) with vanishing conductivity and permittivity. 
 
Into this idealized plasma chamber an arbitrarily shaped probe is 
immersed. The probe contains a finite number of powered electrodes 
$\set E_n$, $n=1\dots N$, which are insulated from each other and
from ground. The electrodes are driven by rf voltages $U_n$.
Grounded surfaces can be treated as another electrode $\set E_0$.
A possible dielectric shielding of the probe is represented as
a part of the subdomain $\set D$ within the plasma chamber $\set V$.

Within the subdomain $\set P$, the dynamical behavior of the plasma, 
given by the dynamics of the charge density $\rho_{\rm e}$ and the 
current density $\vec j_{\rm e}$, is appropriately described by the
cold plasma model in electrostatic approximation. Assuming a complete
electron depletion within the sheath $\set S$, a surface charge
density $\sigma_{\rm e}$ at the sheath edge $\set{K}$ has to be
taken into account. The corresponding equations, including the
constant plasma frequency $\omega_{\rm pe}$ and the collision frequency
for electron-neutral-collisions $\nu$ is given by
\begin{eqnarray}
\frac{\p\sigma_{\rm e}}{\p t} 
& = & \left.-\vec n\cdot\vec j_{\rm e}\frac{}{}\right|_{\vec r \ 
\in\ \set {K}}\ ,
      \nonumber\\[1ex]
\frac{\p\rho_{\rm e}}{\p t} & = & -\di\vec j_{\rm e} 
\label{dynfluid}\ ,\\[1ex]\nonumber
\frac{\p\vec j_{\rm e}}{\p t} 
& = & -\epsi_0\omega_{\rm pe}^2\n\phi-\nu\vec j_{\rm e}
      -\epsi_0\omega_{\rm pe}^2\sum_{n=1}^N U_n\n\psi_n \ . 
\end{eqnarray}
$\phi$ is the inner electrostatic potential and is governed by
Poisson's equation in the domain $\set V$ subject to homogeneous
boundary conditions
\begin{equation}
-\di(\epsi_0 \epsi_r\n\phi)=\left\{\ 
\begin{matrix} 0 & & \vec r & \in & \set D\cup\set S \\[1ex] 
        \sigma_{\rm e} & & \vec r & \in & \set{K} \\[1ex] 
          \rho_{\rm e} & & \vec r & \in & \set P \end{matrix}
	\right. \ .
\label{fluidPoissonIn}          
\end{equation}
The functions $\psi_n$, representing the vacuum coupling between the electrodes, 
are solutions to the homogeneous Poisson equation, 
\begin{equation}
-\di(\epsi_0 \epsi_r\n\psi_n)=0 \ .
\end{equation}
They contain information about the geometry and satisfy the boundary conditions $\psi_n=\delta_{nn'}$ at the electrodes $\set E_n$, with $\delta_{nn'}$ being Kronecker's delta. The respective permittivity $\epsi_r$ is given as $1$ within the
$\set S$ and $\set P$ and with $\epsi_D=\text{const}$ within $\set D$.

The dynamical equations \eqref{dynfluid} can be written in matrix form. This allows to interpret $\sigma_{\rm e}$, $\rho_{\rm e}$, and $\vec j_{\rm e}$ as variables of the state vector $\ket z$, given by  
\begin{equation}
\ket{z} = \vecthree{\sigma_{\rm e}}{\rho_{\rm e}}{\vec j_{\rm e}} \ .%\hvecthree{\sigma_{\rm e}\ ,}{\ \rho_{\rm e}\ ,}{\ \vec j_{\rm e}} \ .
\end{equation}
The state vector is an element of the linear vector space $\set H$. 
For two different state vectors $\ket z$ and $\ket{z'}$ a scalar product, 
which is motivated by the inner energy, can be defined by 
\begin{equation}
\braket{z'}{z}=
\int_{\set V}\epsi\n{\phi '}^{*}\cdot\n\phi\,d^3r
+\int_{\set P}\frac{1}{\epsi_0\omega_{\rm pe}^2}{\vec{j}_{\rm e}^{\ '}}^{*}
 \cdot\vec{j}_{\rm e}\,d^3r\ .
\label{skalarfluid}
\end{equation}
It is compatible with the dynamical equations and induces the corresponding 
norm\linebreak $||z||=\sqrt{\braket{z}{z}^2}$. By means of $||z||$ it is possible to show 
that $\set H$ is complete for all square integrable state vectors including 
singular functions like the surface charge density. Thus, $\set H$ is a Hilbert space.
 
Another important vector is the excitation vector
\begin{equation}
\ket{e_n} = \vecthree{0}{0}{-\epsi_0\omega_{\rm pe}^2\n\psi_n}\ . 
%{\hvecthree{0\ ,}{\ 0\ ,}{\ -\epsi_0\omega_{\rm pe}^2\n\psi_n}}\ .
\end{equation} 
Computing the scalar product between the excitation vector and the state vector, it turns out that this scalar product is equal to the inner current $i_n$ at the electrode $\set E_n$. The inner current represents the observable response of the dynamic system and is given by
\begin{equation}
i_n =\braket{e_n}{z}=-\int_{\set V} \n\psi_n\cdot
\vec j_{\rm e}\, d^3 r \label{fluidStromIn} \ .
\end{equation}
This result shows, that the excitation vector acts also as observation vector. 

Furthermore, two operators can be identified: The conservative
operator $\op T_C$ which is anti-hermitian and the dissipative
operator $\op T_D$ which is hermitian and positive definite. These
operators contain information about the frequency and collisional
damping behavior of the system, respectively. One can find
\begin{eqnarray}
\op{T}_C\ket{z} 
& = & \vecthree{\left. -\vec n\cdot\vec j_{\rm e}\right|_{\vec r\ \in\ 
\set{K}}}{-\di\vec j_{\rm e}}{-\epsi_0\omega_{\rm pe}^2\n\phi}\ ,\\[2ex]
\op{T}_D\ket{z} & = & \vecthree{0}{0}{-\nu\vec j_{\rm e}} \ .
\end{eqnarray}

By means of these definitions it is possible to describe the dynamical
behavior of the probe-plasma system in an abstract, but very compact form
\begin{equation}
\frac{\p}{\p t}\ket z =\op{T}_C\ket{z} + \op{T}_D\ket{z} + 
\sum_{n=1}^N U_n\ket{e_n}
\label{fluidZustGl}\ .
\end{equation}
Concerning measurements, the stationary solutions lie on the focus of interest. Therefore, a harmonic ansatz with the frequency $\omega_{\rm RF}$ is adequate to solve the dynamic equation for the state vector,
\begin{equation}
\ket z =\sum_{n=1}^N\frac{U_n}{i\omega_{\rm RF}-\op{T}_C-\op{T}_D} \ket{e_n}\ .
\label{allgeZustFluid}
\end{equation}
Entering the general solution of the state vector \eqref{allgeZustFluid} into the expression \eqref{fluidStromIn}, one finds that the current $i_n$ is given by the resolvent of the complete dynamical operator $\op T_C+\op T_D$
\begin{equation}
i_n=\braket{e_n}{z}
   =\sum_{n'=1}^N\bra{e_n}\frac{1}{i\omega_{\rm RF}-\op{T}_C-\op{T}_D} \ket{e_{n'}}U_{n'}
   =\sum_{n'=1}^N Y_{nn'}U_{n'}\ .
\label{strom3}
\end{equation}
Thus, the scalar product between two excitation vectors and the resolvent
can be interpreted as the admittance $Y_{nn'}$ between two electrodes,
\begin{equation}
Y_{nn'}=\bra{e_n}\frac{1}{i\omega_{\rm RF}-\op{T}_C-\op{T}_D} \ket{e_{n'}}\ .
\label{VerAdmittanzFluid}
\end{equation}
\pagebreak

The interpretation of equation \eqref{VerAdmittanzFluid} is the main result of the analysis of the general model in ref. \cite{lapke2013}. Based on functional analytic methods, this result can be used to determine an approximated or analytic expression for the admittance between two arbitrary electrodes, which is not derived yet. For this purpose, a complete orthonormal basis $\{\ket{k}\}$ of the Hilbert space is needed. Two of these basis vectors are orthonormal to each other and they satisfy the completeness relation
\begin{equation}
\braket{k'}{k}=\delta_{kk'}\quad ,\quad \sum_{k}\ket{k}\bra{k}=1
\label{OrthoComp}\ .
\end{equation} 
Inserting these expressions into \eqref{VerAdmittanzFluid} allows to expand the admittance $Y_{nn'}$ via the\linebreak orthonormal basis and yields
\begin{equation}
Y_{nn'}%=\bra{e_n}\sum_{k'}\ket{z_{k'}}\bra{z_{k'}}
%\frac{1}{i\omega\op{I}-\op{T}_C-\op{T}_D} \sum_{k}\ket{z_k}\braket{z_{k}}{e_{n'}}
=\sum_{k'}\braket{e_n}{k'}\sum_{k}\bra{k'}
\frac{1}{i\omega_{\rm RF}-\op{T}_C-\op{T}_D} \ket{k}\braket{k}{e_{n'}}\ .
\label{VerAdmittanzFluidExp}
\end{equation}
One can see that the scalar product between two basis vectors and the resolvent
represents the matrix elements of the resolvent's matrix representation.
Based on \eqref{OrthoComp} it is possible to show, that the matrix representation of the resolvent is equal to the inverse matrix representation of the operator $i\omega_{\rm RF}-\op T_C-\op T_D$ 
\begin{equation}
\sum_{k}\bra{k'} i\omega_{\rm RF}-\op{T}_C-\op{T}_D\ket{k}
\bra{k}\frac{1}{i\omega_{\rm RF}-\op{T}_C-\op{T}_D} \ket{k'}=1
\label{OpGleichRes}\ .
\end{equation} 
This result provides the opportunity to first determine the matrix representation of the operator and then to calculate its inverse to find the matrix representation of the resolvent. 

Now, the solution strategy is obvious: One has to choose a set of orthonormal basis functions, after that the matrix elements of the operator $i\omega_{\rm RF}-\op{T}_C-\op{T}_D$ can be determined to find the matrix of the resolvent, then the scalar products between the basis vectors and the excitation vectors have to be computed, and finally the admittance is given by a vector-matrix-vector multiplication.   

In a complex geometry an appropriate set of orthonormal basis functions has to be found to determine an approximated matrix representation of the operators. This leads to an efficient calculation of an approximated admittance instead of a simulation. It can be used to determine the spectral response, e.g., of the plasma absorption probe \cite{sugai1999}. However, in this manuscript we focus on probes with a spherical probe tip. This allows, as we will show in the following sections, to derive an analytic solution of the admittance in a spherical probe-plasma system.

%%%%%%%%%%%%%%%%%%%%%%%%%%%%%%%%%%%%%%%%%%%%%%%%%%%%%%%%%%%%%%%%%%%%%%%%%%%%%%%%
%%%%%%%%%% Eigenvectors of the conservative operator %%%%%%%%%%
%%%%%%%%%%%%%%%%%%%%%%%%%%%%%%%%%%%%%%%%%%%%%%%%%%%%%%%%%%%%%%%%%%%%%%%%%%%%%%%%

\section{Orthonormal basis in spherical geometry}\label{sec:EigenTC}

As shown in the end of the last section, an orthonormal basis is needed to calculate the matrix representation of the resolvent. The ideal basis would be the eigenfunction set of the complete dynamic operator $\op T_C+\op T_D$. However, it is worth noting, that $\op T_C$ and $\op T_D$ do not commute, which means that they do not have the same set of eigenvectors. Therefore, we follow the perturbation approach for operators since the collision frequency $\nu$ in a low pressure plasma is much smaller than the frequency range of interest. 

\begin{figure}[h!]
\psfrag{electrode (E)}{{\tiny electrode ($\set E_1$)}}
\psfrag{sheath (S)}{{\tiny sheath ($\set S$)}}
\psfrag{isolator (I)}{{\tiny isolator ($\set I$)}}
\psfrag{dielectric (D)}{{\tiny dielectric ($\set D$)}}
\psfrag{plasma (P)}{{\tiny plasma ($\set P$)}}
\psfrag{R}{{\tiny $R$}}
\psfrag{R-d}{{\tiny $R-d$}}
\psfrag{R+d1}{{\tiny $R+\delta$}}
\includegraphics[height=5cm]{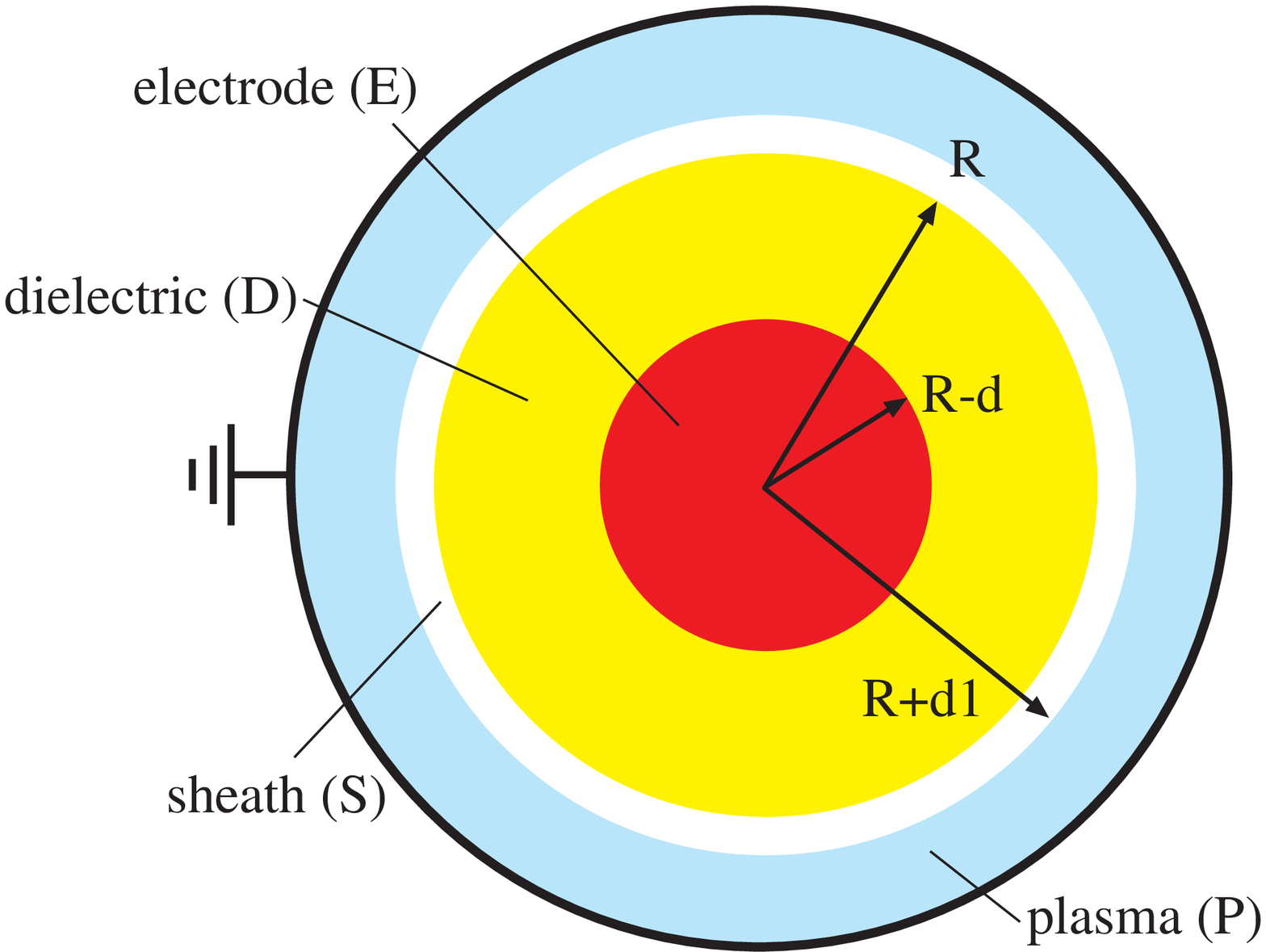}
\hspace{5mm}
\psfrag{electrodes (E)}{\hspace{-3mm}{\tiny electrodes ($\set E_{1/2}$)}}
\includegraphics[height=5cm]{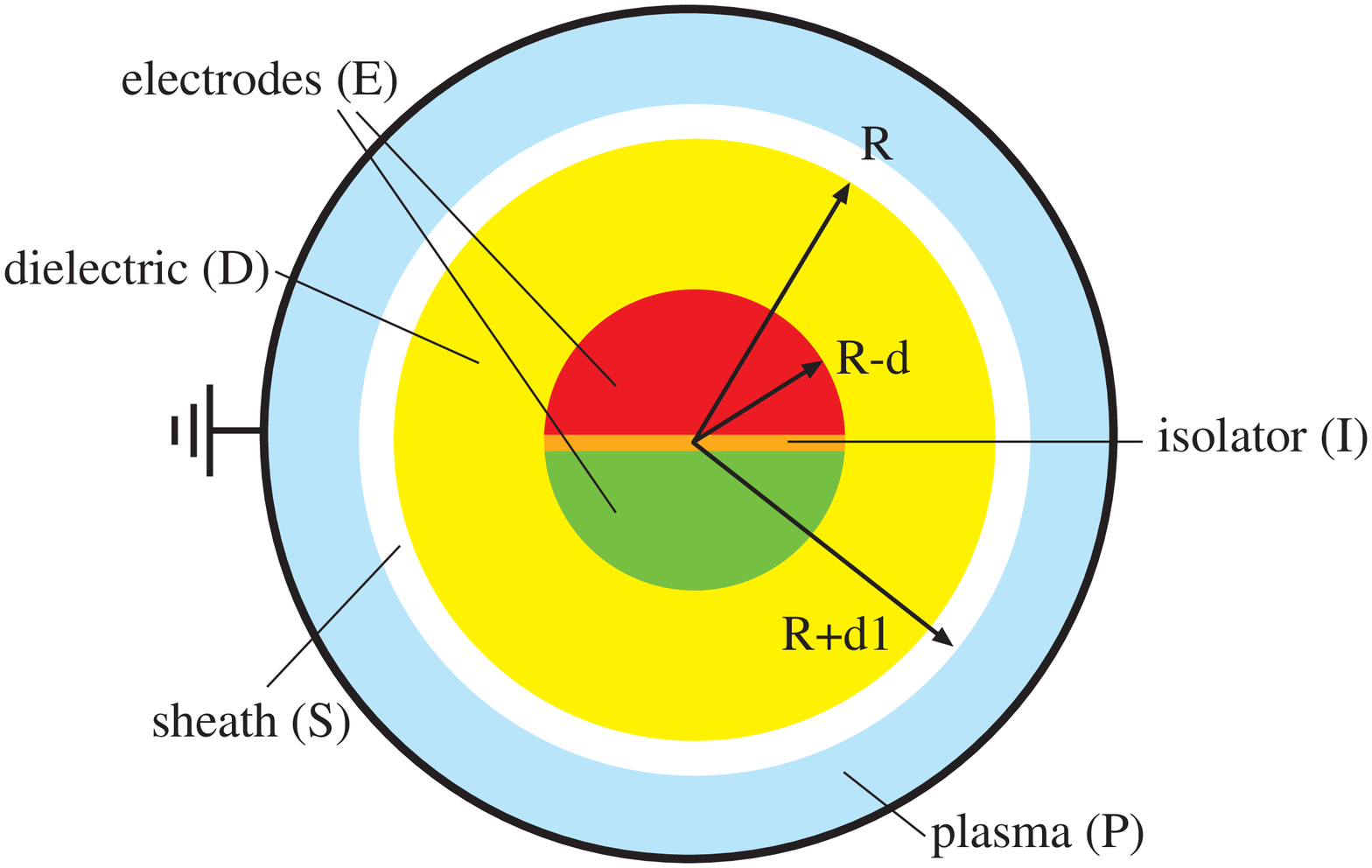}
\caption{Idealized impedance probe (left) and multipole resonance probe (right) with the probe radius $R$, thickness of the dielectric $d$ and sheath thickness $\delta$.}
\label{idealProbes}
\end{figure}

For this purpose we have to determine the eigenvectors of the conservative operator $\op T_C$. Here, we focus on spherical geometry because the idealized spherical impedance probe and the idealized multipole resonance probe have a perfectly spherical geometry. 
They are depicted in figure \ref{idealProbes} with the probe radius $R$, the thickness of the dielectric $d$, and the sheath thickness $\delta$. 
  
Due to the fact that $\op T_C$ is anti-hermitian the eigenvalue 
equation can be written with a pure imaginary eigenvalue
%\begin{equation}
$i\omega\ket z=\op T_C\ket z\ .$
%\label{EigenGlTc}
%\end{equation}
To solve the eigenvalue problem in spherical geometry we expand all 
scalar functions in spherical harmonics 
\begin{eqnarray}
\sigma_{\rm e}(\vartheta,\varphi) & = & 
\sum_{l=0}^\infty \sum_{m=-l}^l\sigma_{lm}Y_{lm}(\vartheta,\varphi)\ , \nonumber\\
\rho_{\rm e}(r,\vartheta,\varphi) & = & 
\sum_{l=0}^\infty \sum_{m=-l}^l\rho_{lm}(r)Y_{lm}(\vartheta,\varphi)\ ,\\
\phi(r,\vartheta,\varphi)   & = & 
\sum_{l=0}^\infty \sum_{m=-l}^l\phi_{lm}(r)Y_{lm}(\vartheta,\varphi)\ ,\nonumber
\nonumber
\end{eqnarray}
and the current density in vectorial spherical harmonics
\begin{equation}
\vec j_{\rm e}(r,\vartheta,\varphi)=
\sum_{l=0}^\infty \sum_{m=-l}^l 
 j_{lm}^{\,(X)}(r)\vec X_{lm}(\vartheta,\varphi)
+j_{lm}^{\,(Y)}(r)\vec Y_{lm}(\vartheta,\varphi)
+j_{lm}^{\,(Z)}(r)\vec Z_{lm}(\vartheta,\varphi)\ .
\end{equation}
This expansion leads to the following set of equations:
\begin{eqnarray}
i\omega\sigma_{lm} & = & \left.-j_{lm}^{\,(Y)}\right|_{r=R+\delta}\ ,
\label{EigenGlSigma}\\
i\omega\rho_{lm}   
& = & -\left[\frac{1}{r^2}\frac{\p}{\p r}\left(r^2 j_{lm}^{\,(Y)}\right)
      -\frac{i}{r}\sqrt{l(l+1)} j_{lm}^{\,(Z)}\right]\ ,\label{EigenGlRho}\\
i\omega j_{lm}^{\,(X)} & = & 0\ ,\label{EigenGlJx}\\
i\omega j_{lm}^{\,(Y)} 
& = & -\epsi_0\omega_{\rm pe}^2\frac{\p}{\p r}\phi_{lm}\ ,\label{EigenGlJy}\\
i\omega j_{lm}^{\,(Z)} 
& = & \epsi_0\omega_{\rm pe}^2\frac{i}{r}\sqrt{l(l+1)}\phi_{lm}\label{EigenGlJz}\ .
\end{eqnarray}
Consequently, Poisson's equation reads
\begin{equation}
-\epsi_0\left[
 \frac{1}{r^2}\frac{\p}{\p r}\left(r^2\frac{\p}{\p r}\phi_{lm}\right)
 -\frac{l(l+1)}{r^2}\phi_{lm}\right]
= \left\{\ 
\begin{matrix}    0 & & r & \in & [R-d,R+\delta)\\[1ex] 
        \sigma_{lm} & & r &  =  & R+\delta \\[1ex] 
          \rho_{lm} & & r & \in & [R+\delta,\infty) \end{matrix}\right. 
\label{fluidPoissonInExp}
\end{equation}
with the corresponding boundary conditions
\begin{equation}
\phi_{lm}^{(\set D)}(R-d) = 0 \quad \text{and}
\quad
\lim_{r\ra\infty}\phi_{lm}^{(\set P)}(r)  =  0 \label{RB}\ ,
\end{equation}
and transition conditions
\begin{gather}
\begin{matrix}
\phi_{lm}^{(\set D)}(R)-\phi_{lm}^{(\set S)}(R) & = & 0 & , &
\phi_{lm}^{(\set P)}(R+\delta)-\phi_{lm}^{(\set S)}(R+\delta) & = & 0 \ ,\\[2ex]
\ds\left.\left(\varepsilon_D\frac{\p\phi_{lm}^{(\set D)}}{\p r}\right.
   -\left.\frac{\p\phi_{lm}^{(\set S)}}{\p r}\right)\right|_{r=R} 
& = & 0 & , & -\epsi_0 \ds \left.\left(\frac{\p\phi_{lm}^{(\set P)}}{\p r} -\frac{\p\phi_{lm}^{(\set S)}}{\p r}\right)\right|_{r=R+\delta} & = & \sigma_{lm} \ . 
\end{matrix}
\label{TB}
\end{gather}\\[-5mm]

From \eqref{EigenGlSigma} to \eqref{EigenGlJz} it is obvious that two cases have to be distinguished: $\omega=0$ and $\omega\neq0$. Indeed $\omega=0$ is an eigenvalue of the operator $\op T_C$, but it is not excited by the harmonic RF voltages applied to the electrodes of the probe. Due to that it has no contribution to the response function and will not be considered in the rest of the manuscript. The case $\omega\neq0$ is more important. Combining the dynamical equations \eqref{EigenGlRho}, \eqref{EigenGlJy}, and \eqref{EigenGlJz} with Poisson's equation \eqref{fluidPoissonInExp} one obtains
\begin{equation}
\left(1-\frac{\omega_{\rm pe}^2}{\omega^2}\right)
\left[\frac{1}{r^2}\frac{\p}{\p r}\left(r^2\frac{\p}{\p r}\phi_{lm}^{(\set P)}\right)
     -\frac{l(l+1)}{r^2}\phi_{lm}^{(\set P)}\right]=0\ .
\label{LaplacePlasma}
\end{equation}
And again, two cases can be distinguished: $\omega=\pm\omega_{\rm pe}$ 
and $\omega\neq\pm\omega_{\rm pe}$.\\

%%%%%%%%%%%%%%%%%%%%%%%%%%%%%%%%%%%%%%%%%%%%%%%%%%%%%%%%%%%%%%%%%%%%%%%%%%%%%%%%
%%%%%%%%%% Eigenfunktionen f�r omega ungleich omega_p %%%%%%%%%%
%%%%%%%%%%%%%%%%%%%%%%%%%%%%%%%%%%%%%%%%%%%%%%%%%%%%%%%%%%%%%%%%%%%%%%%%%%%%%%%%

Since for $\omega\neq\pm\omega_{\rm pe}$, the potential in the plasma has to 
satisfy Laplace's equation. The same holds of course for the potential 
in $\set D$ and $\set S$. Thus, all potentials are governed by the same 
equation because the permittivity is constant in the different regions
\begin{equation}
\frac{1}{r^2}\frac{\p}{\p r}
\left(r^2\frac{\p}{\p r}\phi^{(\set D, \set S, \set P)}_{lm}\right)
        -\frac{l(l+1)}{r^2}\phi^{(\set D, \set S, \set P)}_{lm}=0 \ .
\end{equation}
Its general solution is given by 
\begin{equation}
\phi_{lm}^{(\set D, \set S, \set P)}(r)
=A^{(\set D, \set S, \set P)}r^l
+B^{(\set D, \set S, \set P)}r^{-(l+1)}\ .
\label{allgeLaplaceLsg}
\end{equation}

The coefficients $A^{(\set S, \set P)}$ and $B^{(\set D, \set S, \set P)}$ %-- they are explicitly given in the appendix \eqref{coeff} -- 
are determined by five of the boundary and transition conditions given in \eqref{RB} and \eqref{TB} 
\begin{eqnarray}
B^{(\set D)} & = & -\frac{A^{(\set D)}}{(R-d)^{2l+1}}
               =:   -\frac{A_{lm}}{(R-d)^{2l+1}}\nonumber\\[1ex]
A^{(\set S)} & = & \frac{A^{(\set D)}}{2l+1}
                   \left[1+l(\epsi_D+1)+(l+1)(\epsi_D-1)
                   \left(1-\frac{d}{R}\right)^{2l+1}\right]
                   =:A_{lm}A^{(\set S)}_l\nonumber\\[1ex]
B^{(\set S)} & = & \frac{A^{(\set D)}}{2l+1}
                   \left[l(1-\epsi_D)R^{2l+1}-(l+\epsi_D(l+1))(R-d)^{2l+1}\right]
                   =:A_{lm}B^{(\set S)}_l\label{coeff}\\[1ex]
A^{(\set P)} & = & 0\nonumber\\[1ex]
B^{(\set P)} & = & \frac{A^{(\set D)}}{2l+1}
                   \left\{l(1-\epsi_D)R^{2l+1}-(l+(l+1)\epsi_D)(R-d)^{2l+1}
                   \frac{}{}\right.\nonumber\\[1ex]
             &   & +\left.(R+\delta)^{2l+1}
                   \left[(l+1)(\epsi_D-1)
                   \left(1-\frac{d}{R}\right)^{2l+1}+1+l(1+\epsi_D)\right]\right\}
                   =:A_{lm}B^{(\set P)}_l\nonumber
\end{eqnarray}
Each of these coefficients depends on $A^{(\set D)}=A_{lm}$ and they allow for the complete inner potential to be written as
\begin{equation}
\phi(\vec r)%=\sum_{l,m}A_{lm}\phi_{lm} Y_{lm} 
=\sum_{l,m}A_{lm}Y_{lm}\left\{
\begin{matrix} 
\ds r^l-(R-d)^{2l+1}r^{-(l+1)} & , & \vec r & \in & \set D\\[1ex]
\ds A^{(\set S)}_lr^l
   +B^{(\set S)}_lr^{-(l+1)} & , & \vec r & \in & \set S\\[1ex]
\ds B^{(\set P)}_lr^{-(l+1)} & , & \vec r & \in & \set P
\end{matrix}\right.\ .
\end{equation}
The sixth condition given in \eqref{RB} and \eqref{TB} 
determines the eigenvalues of $\op T_C$ in 
spherical coordinates
\begin{eqnarray}
\omega_{lm} 
& = & \pm\omega_{\rm pe}\sqrt{\frac{l+1}{2l+1}
      \left[1-\frac{l(\epsi_D-1)+(\epsi_D(l+1)+l)\left(1-\frac{d}{R}\right)^{2l+1}}
                   {1+l(1+\epsi_D)+(l+1)(\epsi_D-1)\left(1-\frac{d}{R}\right)^{2l+1}}
      \left(1+\frac{\delta}{R}\right)^{-(2l+1)}\right]}\nonumber\\
& =: & \pm\omega_{\rm pe}\eta_l\label{fluidResFreq}\ .
\end{eqnarray}
They are proportional to the electron plasma frequency and thus, simply connected to the electron density. It is important to note that the eigenvalues are independent of the index $m$. This shows that the resonance modes of a spherical probe-plasma system described by the cold plasma model are always symmetric referred to a rotation around an arbitrary rotation axis. The corresponding charge, surface charge, and current density depend on the inner potential defined by the equations \eqref{EigenGlSigma} to \eqref{EigenGlJz}.\\
%\begin{equation}
%\vec j_{\rm e}(\vec r)
% =\sum_{l,m}A_{lm}\epsi_0\frac{\omega_{\rm pe}^2}{\omega_{lm}}
%  \left[-i(l+1)\vec Y_{lm}+\sqrt{l(l+1)}\vec Z_{lm}\right]
%  \frac{B^{(\set P)}_l}{r^{l+2}}\ ,\ \forall\ \vec r\in\set P \ .
%\end{equation}
%%%%%%%%%%%%%%%%%%%%%%%%%%%%%%%%%%%%%%%%%%%%%%%%%%%%%%%%%%%%%%%%%%%%%%%%%%%%%%%%
%%%%%%%%%% Eigenfunktionen f�r omega gleich omega_p %%%%%%%%%%
%%%%%%%%%%%%%%%%%%%%%%%%%%%%%%%%%%%%%%%%%%%%%%%%%%%%%%%%%%%%%%%%%%%%%%%%%%%%%%%%

In the case $\omega=\pm\omega_{\rm pe}$ an arbitrary potential fulfills equation \eqref{LaplacePlasma} in the plasma region 
\begin{equation}
\phi_{lm}^{(\set P)}(r)=\text{arbitrary}\ ,\ \forall \vec r\in\set P\quad \text{with} \quad \lim_{r\ra\infty}\phi_{lm}^{(\set P)}(r)=0\ .
\end{equation}
The potential in $\set S$ and $\set D$ vanishes due to the rewritten 
boundary and transition conditions in this case and the complete 
inner potential is given by
\begin{equation}
\phi(\vec r)=\sum_{l,m}Y_{lm}\left\{
\begin{matrix} 
0 & , & \vec r & \in & \set S\cup\set D\\
\phi_{lm}^{(\set P)}(r) & , & r & \in & \set P
\end{matrix}\right.\ .
\end{equation}
Again, the corresponding charge, surface charge, and current density depend on the inner potential. \\
%The corresponding current density is determined by
%\begin{equation}
%\vec j_{\rm e}(\vec r)= \epsi_0\omega_{\rm pe}\sum_{l,m}
% \left[i\frac{\p}{\p r}\phi_{lm}^{(\set P)}(r)\vec Y_{lm}
%       +\frac{\sqrt{l(l+1)}}{r}\phi_{lm}^{(\set P)}(r)\vec Z_{lm}\right]
% \ ,\ \forall\ \vec r\in\set P\ .      
%\end{equation}\\[-5mm]

In summary, we find the following two different eigenvectors of the 
conservative operator $\op T_C$, which build a complete orthogonal 
set in the Hilbert space
\begin{description}
\item[$\omega=\pm\omega_{lm}:$] $\quad$\\%[-8mm]
\begin{equation}
%\begin{eqnarray}
%\omega=\pm\omega_{lm}:\ 
\ket{z_{lm}^{(1\pm)}} =
\begin{pmatrix} 
\ds\epsi_0\frac{\omega_{\rm pe}^2}{\omega_{lm}^2} \frac{l+1}{(R+\delta)^{l+2}}Y_{lm}\\[3ex]
0\\[1ex]
\ds\pm\epsi_0\frac{\omega_{\rm pe}^2}{\omega_{lm}}
\left[-i(l+1)\vec Y_{lm}+\sqrt{l(l+1)}\vec Z_{lm}\right]
\frac{B^{(\set P)}_l}{r^{l+2}}
\end{pmatrix}\\[1cm]
\end{equation}
\end{description}
%\item[$\omega=0:$] 
%\begin{equation}
%\ket{z_{lm}^{(2)}}=
%\begin{pmatrix} 
%\ds 0\cdot Y_{lm}\\[1ex]
%\ds j_{lm}^{(X)}(r)\vec X_{lm}+j_{lm}^{(Y)}(r)\vec Y_{lm}+j_{lm}^{(Z)}(r)\vec Z_{lm}
%\end{pmatrix}
%\end{equation}
\begin{description}
\item[$\omega=\pm\omega_{\rm pe}:$] $\quad$\\%[-8mm]
\begin{equation}
%\omega=\pm\omega_{\rm pe}:\ 
\ket{z_{lm}^{(2\pm)}} = 
\begin{pmatrix}  
\ds -\epsi_0\left.\frac{\p}{\p r}\phi_{lm}^{(\set P)}\right|_{r=R+\delta}Y_{lm}\\[4ex]
\ds -\frac{\epsi_0}{r^2}\left[
     \frac{\p}{\p r}\left(\frac{\p}{\p r}\phi_{lm}^{(\set P)}\right) 
    -l(l+1)\phi_{lm}^{(\set P)}\right]Y_{lm}\\[4ex]
\ds\pm\epsi_0\omega_{\rm pe}\left[i\frac{\p}{\p r}\phi_{lm}^{(\set P)}(r)\vec Y_{lm}
 +\frac{\sqrt{l(l+1)}}{r}\phi_{lm}^{(\set P)}(r)\vec Z_{lm}\right]
\end{pmatrix}%\qquad
%\end{eqnarray}
\end{equation}\\[5mm]
\end{description}
%It is important to note that the inner potential $\phi=\phi\{\rho_{\rm e},\sigma_{\rm e}\}$ is a linear functional of the charge density $\rho_{\rm e}$ and the surface charge
%density $\sigma_{\rm e}$. They are uniquely coupled to each other via Poisson's equation. This allows to introduce the inner potential in the eigenvectors instead of the surface charge and the charge density. \\

A scalar product naturally induces a norm, whereby the eigenvectors can be normalized. In spherical geometry it is possible to simplify the scalar product \eqref{skalarfluid} via the expansion in the orthogonal spherical harmonics 
\begin{eqnarray}
\braket{z'}{z}
& = & \sum_{l,m}\int_{R-d}^{\infty} 
      \epsi\left(\left|\frac{\p\phi_{lm}}{\p r}\right|^2
     +\frac{l(l+1)}{r^2}\left|\phi_{lm}\right|^2
      \right)r^2\,dr\nonumber\\
&   &+\frac{1}{\epsi_0\omega_{\rm pe}^2}\sum_{l,m}\int_{R+\delta}^{\infty}
      \left(\left| j_{lm}^{(X)}\right|^2 +\left| j_{lm}^{(Y)}\right|^2 
     +\left| j_{lm}^{(Z)}\right|^2 \right)r^2\,dr \ .
\label{skalarSimp}    
\end{eqnarray}
By means of the simplified scalar product \eqref{skalarSimp} each norm of the different eigenvectors can be determined. Introducing the components of the inner potential and
the current density into the scalar product, the squared norm of the 
first eigenvector can explicitly be evaluated 
\begin{equation}
%\begin{eqnarray}
 \left|\left|z_{lm}^{(1)}\right|\right|^2
 = \braket{z_{lm}^{(1)}}{z_{lm}^{(1)}}
  =  \frac{2\epsi_0(l+1)B_l^{(\set P)\, 2}}{(R+\delta)^{2l+1}\eta_l^2}\ .
\end{equation}
%In the second eigenvector the inner potential is equal to zero. 
%The corresponding norm obtains only the current components, which are 
%arbitrary, and is given by
%\begin{equation}
%\left|\left|z_{lm}^{(2)}\right|\right|^2
%=\braket{z_{lm}^{(2)}}{z_{lm}^{(2)}}
%=\frac{1}{\epsi_0\omega_{\rm pe}^2}\int_{R+\delta}^{\infty}
% \left(|j_{lm}^{(X)}|^2+|j_{lm}^{(Y)}|^2+|j_{lm}^{(Z)}|^2\right)r^2\,dr\ 
%\end{equation}
The elements of the second eigenvector depend on the inner potential $\phi_{lm}^{(\set P)}$ in the plasma region $\set P$. It turns out that the remaining integrals in the scalar product are equal and the norm results in
\begin{equation}
\left|\left|z_{lm}^{(2)}\right|\right|^2=\braket{z_{lm}^{(2)}}{z_{lm}^{(2)}}
= 2\epsi_0\int_{R+\delta}^{\infty}
   \left[\left|\frac{\p\phi_{lm}^{(\set P)}}{\p r}\right|^2
  +\frac{l(l+1)}{r^2}\left|\phi_{lm}^{(\set P)}\right|^2\right]r^2\,dr \ .
\end{equation}
The integral remains undetermined due to the arbitrary inner potential in $\set P$.\\

Once the norms are determined, we are able to define the following completeness 
relation because the normalized eigenvectors build a complete orthonormal set in the Hilbert space
\begin{align}
\sum_{lm}\left(\ket{\hat{z}_{lm}^{(1+)}}\bra{\hat{z}_{lm}^{(1+)}}%\right.
%&
+\ket{\hat{z}_{lm}^{(1-)}}\bra{\hat{z}_{lm}^{(1-)}}
 %+\ket{\hat{z}_{lm}^{(2)}}\bra{\hat{z}_{lm}^{(2)}}
% \nonumber\\
%&
+%\left. 
\ket{\hat{z}_{lm}^{(2+)}}\bra{\hat{z}_{lm}^{(2+)}}
 +\ket{\hat{z}_{lm}^{(2-)}}\bra{\hat{z}_{lm}^{(2-)}}\right)=1\ .
\label{vollrelkugel}                  
\end{align}
As seen in equation \eqref{VerAdmittanzFluidExp} the completeness relation is needed for the expansion and simplification of the admittance given in equation \eqref{VerAdmittanzFluid}. Additionally, it allows the expansion of an arbitrary state vector in the eigenvectors of the conservative operator 
\begin{equation}
\ket{z}=\sum_{l,m}
 A_{lm}^{(1+)}\ket{\hat z_{lm}^{(1+)}}%}{\ \left|\left| z_{lm}^{(1)}\right|\right|}
+A_{lm}^{(1-)}\ket{\hat z_{lm}^{(1-)}}%}{\ \left|\left| z_{lm}^{(1)}\right|\right|}
%+A_{lm}^{(2)}\frac{\ket{z_{lm}^{(2)}}}{\ \left|\left| z_{lm}^{(2)}\right|\right|}
+A_{lm}^{(2+)}\ket{\hat z_{lm}^{(2+)}}%}{\ \left|\left| z_{lm}^{(2)}\right|\right|}
+A_{lm}^{(2-)}\ket{\hat z_{lm}^{(2-)}}%}{\ \left|\left| z_{lm}^{(2)}\right|\right|}
\ .
\end{equation}\\

%%%%%%%%%%%%%%%%%%%%%%%%%%%%%%%%%%%%%%%%%%%%%%%%%%%%%%%%%%%%%%%%%%%%%%%%%%%%%%%%
%%%%%%%%%% General excitation state vectors %%%%%%%%%%
%%%%%%%%%%%%%%%%%%%%%%%%%%%%%%%%%%%%%%%%%%%%%%%%%%%%%%%%%%%%%%%%%%%%%%%%%%%%%%%%

\section{General excitation vector}

After the orthonormal basis is derived, we have to compute the excitation vector to determine the admittance \eqref{VerAdmittanzFluidExp}. Based on the specified geometry we are able to calculate the general excitation vector $\ket{e_n}$. It contains the 
characteristic functions $\psi_n$, which follow Laplace's equation 
\begin{equation}
\n\cdot(\epsi_0\epsi_r\n\psi_n)=0\qquad \text{with}\quad \left.\lim_{r\ra\infty}\psi_n=0\quad \text{and}\quad 
\psi_n\right|_{\set E_{n'}}=\delta_{nn'}\ .
\end{equation}
Similar to the eigenvector calculation we expand
the characteristic functions in spherical harmonics and 
determine the solution in $r$-direction 
\begin{equation}
\psi_{lm}(r) 
 = \left\{
\begin{matrix} 
\ds \alpha^{(\set D)} r^l+\beta^{(\set D)} r^{-(l+1)} & , & r & \in & [R-d,R)\\[2ex]
\ds \beta^{({\rm vac})} r^{-(l+1)} & , & r & \in & [R,\infty)
\end{matrix}\right.\ .
\end{equation}
Due to the continuity of the vacuum potential and the 
electric flux density at the surface of the dielectric 
we find the following transition conditions
\begin{equation}
\psi_{lm}^{(\set D)}(R)=\psi_{lm}^{({\rm vac})}(R)\qkq
\epsi_D\left.\frac{\p}{\p r}\psi_{lm}^{(\set D)}\right|_R 
=\left.\frac{\p}{\p r}\psi_{lm}^{({\rm vac})}\right|_R\ .
\end{equation}
They allow for $\alpha^{(\set D)}$ and $\beta^{(\set D)}$ to be determined dependent on 
$\beta^{({\rm vac})}=\beta_{lm}^{(n)}$ 
\begin{equation}
\alpha^{(\set D)}=\frac{(l+1)(\epsi_D-1)}{(2l+1)\epsi_D R^{2l+1}}\beta^{(\rm vac)}\qkq
\beta^{(\set D)}=\frac{1+l(1+\epsi_D)}{(2l+1)\epsi_D}\beta^{(\rm vac)}\ .
\end{equation}
By means of these coefficients the general characteristic functions are defined as
\begin{equation}
\psi_n(\vec r)
=\sum_{l,m}\beta_{lm}^{(n)}\psi_{lm}^{(n)}(r)Y_{lm}(\vartheta, \varphi)
\label{charFkt}
\end{equation}
and determine the general excitation vector to
\begin{equation}
\ket{e_n}=\sum_{l,m}\ket{0 \ ,\ 0\ ,\ 
\ds-\epsi_0\omega_{\rm pe}^2\beta_{lm}^{(n)}\left(\frac{\p\psi_{lm}^{(n)}}{\p r}\vec Y_{lm}
-\frac{i}{r}\sqrt{l(l+1)}\psi_{lm}^{(n)}\vec Z_{lm}\right)}\ .
\end{equation}
The remaining coefficient $\beta_{lm}^{(n)}$ can be evaluated 
by the boundary condition $\psi_{lm}^{(n)}(R-d)=\delta_{nn'}$ 
at the electrodes $\set E_n$. Utilizing the orthogonality 
of the spherical harmonics we find
\begin{equation}
\beta_{lm}^{(n)} 
=\frac{1}{\gamma_l}\int_{\set E_n}Y_{lm}^*(\vth,\vp)\,d\Omega
\label{FourierKoef}
%\end{equation}
\quad \text{with}\quad
%\begin{equation}
\gamma_{l}
:= \frac{(l+1)(\epsi_D-1)\left(1-\frac{d}{R}\right)^{2l+1}+1+l(1+\epsi_D)}
        {(2l+1)\epsi_D R^{l+1}\left(1-\frac{d}{R}\right)^{l+1}} \ .
\end{equation}
The integral in $\beta_{lm}^{(n)}$ contains the information 
about the electrode configuration within the probe tip and 
has to remain undetermined until the configuration is defined.

%%%%%%%%%%%%%%%%%%%%%%%%%%%%%%%%%%%%%%%%%%%%%%%%%%%%%%%%%%%%%%%%%%%%%%%%%%%%%%%%
%%%%%%%%%% General admittance in spherical geometry %%%%%%%%%%
%%%%%%%%%%%%%%%%%%%%%%%%%%%%%%%%%%%%%%%%%%%%%%%%%%%%%%%%%%%%%%%%%%%%%%%%%%%%%%%%

\section{General admittance in spherical geometry}

Now, we are equipped with all the necessary elements to expand the 
admittance in equation \eqref{VerAdmittanzFluid} via the completeness 
relation \eqref{vollrelkugel} and determine the general admittance 
in a spherical geometry. Introducing the completeness relation \eqref{vollrelkugel} twice 
into \eqref{VerAdmittanzFluid} between the excitation vectors and the resolvent yields a long expression, in which scalar products between the eigenvectors and the excitation 
vectors appear. The scalar products between the excitation state vector and second eigenvector becomes zero and simplifies the admittance
\begin{align}
Y_{nn'} 
= \sum_{lm}\sum_{l'm'}&
  \braket{e_n}{\hat z_{lm}^{(1+)}}\bra{\hat z_{lm}^{(1+)}}
  (i\omega_{\rm RF}-\op T_C-\op T_D)^{-1}\ket{\hat z_{l'm'}^{(+)}}
  \braket{\hat z_{l'm'}^{(1+)}}{e_{n'}}\nonumber\\[1ex]
&+\braket{e_n}{\hat z_{lm}^{(1+)}}\bra{\hat z_{lm}^{(1+)}}
  (i\omega_{\rm RF}-\op T_C-\op T_D)^{-1}\ket{\hat z_{l'm'}^{(1-)}}
  \braket{\hat z_{l'm'}^{(1-)}}{e_{n'}}\label{AdmittanzKugelSimp}\\[2ex]
&+\braket{e_n}{\hat z_{lm}^{(1-)}}\bra{\hat z_{lm}^{(1-)}}
  (i\omega_{\rm RF}-\op T_C-\op T_D)^{-1}\ket{\hat z_{l'm'}^{(1+)}}
  \braket{\hat z_{l'm'}^{(1+)}}{e_{n'}}\nonumber\\[2ex]
&+\braket{e_n}{\hat z_{lm}^{(1-)}}\bra{\hat z_{lm}^{(1-)}}
  (i\omega_{\rm RF}-\op T_C-\op T_D)^{-1}\ket{\hat z_{l'm'}^{(1-)}}
  \braket{\hat z_{l'm'}^{(1-)}}{e_{n'}}\nonumber  \ .
\end{align}
The remaining scalar products between $\ket{e_n}$ and $\ket{\hat z_{lm}^{(1\pm)}}$ 
can explicitly be evaluated. They differ only in their sign and the expansion index. Thus they are given by

\begin{eqnarray}
\braket{e_n}{\hat z_{lm}^{(+)}}
& = & -\frac{i\epsi_0\omega_{\rm pe} (l+1)\beta_{lm}^{(n)}B_l^{(\set P)}}
            {\eta_{l}(R+\delta)^{2l+1}\left|\left|z_{lm}^{(1)}\right|\right|} =%\qkq
-\braket{e_n}{\hat z_{lm}^{(-)}}
%=\frac{i\epsi_0\omega_{\rm pe} (l+1)\beta_{lm}^{(n)}B_l^{(\set P)}}
 %           {\eta_{l}(R+\delta)^{2l+1}\left|\left|z_{lm}^{(1)}\right|\right|}
 \ ,\label{SkalarEigenAnreg1}\\[2ex]
\braket{\hat z_{l'm'}^{(+)}}{e_{n'}}
& = & \frac{i\epsi_0\omega_{\rm pe} (l'+1)\beta_{l'm'}^{(n')}B_{l'}^{(\set P)}}
           {\eta_{l'}(R+\delta)^{2l'+1}\left|\left|z_{l'm'}^{(1)}\right|\right|} =%\qkq
-\braket{\hat z_{l'm'}^{(-)}}{e_{n'}}
%=-\frac{i\epsi_0\omega_{\rm pe} (l'+1)\beta_{l'm'}^{(n')}B_{l'}^{(\set P)}}
%       {\eta_{l'}(R+\delta)^{2l'+1}\left|\left|z_{l'm'}^{(1)}\right|\right|}
      \label{SkalarEigenAnreg2}\ .\\\nonumber
\end{eqnarray}

Finally, we have to evaluate the matrix elements of the resolvent. 
Equation \eqref{OpGleichRes} shows that the matrix of the 
resolvent can be calculated by the inverse of the matrix of the operator
$i\omega_{\rm RF}-\op T_C-\op T_D$. As an example, we compute the matrix element 
concerning to the first term in equation \eqref{AdmittanzKugelSimp}, 
whereas the eigenvalue representation of the conservative operator $\op T_C$ is used
\begin{equation}
\bra{\hat z_{l'm'}^{(1+)}}i\omega_{\rm RF}-\op T_C-\op T_D\ket{\hat z_{lm}^{(1+)}}
=i(\omega_{\rm RF}-\omega_{\rm pe}\eta_{l})\delta_{ll'}\delta_{mm'}
-\bra{\hat z_{l'm'}^{(1+)}}\op T_D\ket{\hat z_{lm}^{(1+)}}\ .
\label{matrixelement}
\end{equation}\\
%\pagebreak

\noindent The matrix elements of $\op T_D$ have to be evaluated explicitly. 
Applying the operator $\op T_D$ to $\ket{\hat z_{lm}^{(1+)}}$ 
yields a state vector with a vanishing charge contribution. 
Thus, the scalar product with $\bra{\hat z_{lm}^{(1+)}}$ is 
reduced to the integral over the current components and is determined by
\begin{align}
\bra{\hat z_{l'm'}^{(1+)}}\op T_D\ket{\hat z_{lm}^{(1+)}}
= & -\frac{\nu\epsi_0\left[(l+1)(l'+1)+\sqrt{l(l+1)l'(l'+1)}\right]}
            {(l+l'+1)\eta_{l}\eta_{l'}\norm{z_{lm}^{(1)}}\norm{z_{l'm'}^{(1)}}(R+\delta)^{l+l'+1}}
       B_l^{(\set P)}B_{l'}^{(\set P)}\delta_{ll'}\delta_{mm'}\nonumber\\
=:&-\nu_{ll'}\delta_{ll'}\delta_{mm'}\label{matrixelementTD}\ .
\end{align}
Entering \eqref{matrixelementTD} in \eqref{matrixelement} yields 
the complete first matrix element
\begin{equation}
\bra{\hat z_{l'm'}^{(+)}}i\omega_{\rm RF}-\op T_C-\op T_D\ket{\hat z_{lm}^{(+)}}
=\left[i(\omega_{\rm RF}-\omega_{\rm pe}
 \eta_{l})+\nu_{ll'}\right]\delta_{ll'}\delta_{mm'}\ .
\label{MatrixElementPlus}
\end{equation}
The fact that $\op T_D$ is represented by pure diagonal elements shows 
that $\op T_C$ and $\op T_D$ can be projected on the same subdomain with spherical harmonics as basisfunctions. Therefore, the perturbation approach becomes 
dispensable. Following the same calculation, we find the other matrix elements 
of the dynamical operator
\begin{eqnarray}
\bra{\hat z_{l'm'}^{(1-)}}\op T_D\ket{\hat z_{lm}^{(1-)}}
& = & \left[i(\omega_{\rm RF}+\omega_{\rm pe}\eta_{l})+\nu_{ll'}\right]
      \delta_{ll'}\delta_{mm'}\\[1ex]
      \bra{\hat z_{l'm'}^{(1+)}}\op T_D\ket{\hat z_{lm}^{(1-)}}
& = & \bra{\hat z_{l'm'}^{(1-)}}\op T_D\ket{\hat z_{lm}^{(1+)}}
  =  -\nu_{ll'}\delta_{ll'}\delta_{mm'}\ .
\end{eqnarray}
This result shows that the operator $i\omega_{\rm RF}-\op T_C-\op T_D$ is represented by 
a diagonal block matrix, where the block elements on the main diagonal are given 
by $(2\times2)$-matrices. The inverse of such a matrix is also a diagonal 
block matrix with $(2\times2)$-matrices on the main diagonal. Each block element on 
the diagonal is determined by the inverse of the $(2\times2)$-matrix 
of the original block element.% and is given by
%\begin{equation}
%\begin{pmatrix}
%i(\omega_{\rm RF}-\omega_{\rm pe}\eta_{l})+\nu_{ll'} & -\nu_{ll'}\\
%-\nu_{ll'} & i(\omega_{\rm RF}+\omega_{\rm pe}\eta_{l})+\nu_{ll'}
%\end{pmatrix}^{-1}
%=\frac{
%\begin{pmatrix}
%i(\omega_{\rm RF}+\omega_{\rm pe}\eta_{l})+\nu_{ll'} & \nu_{ll'}\\
%\nu_{ll'} & i(\omega_{\rm RF}-\omega_{\rm pe}\eta_{l})+\nu_{ll'}
%\end{pmatrix}}{\omega_{\rm pe}^2\eta_{l}^2+2i\omega_{\rm RF}\nu_{ll'}-\omega_{\rm RF}^2}\, .
%\label{DiagonalElement}
%\end{equation}

Now, all terms in the admittance \eqref{AdmittanzKugelSimp} are 
evaluated and can be introduced. Finally, we exploit the Kronecker 
deltas to determine the general admittance in a spherical geometry 
\begin{align}
Y_{nn'} 
=i\omega\epsi_0\omega_{\rm pe}^2\sum_{l=0}^\infty\sum_{m=-l}^l
 \frac{l+1}{(R+\delta)^{2l+1}}
 \frac{\beta_{lm}^{(n)}\beta_{lm}^{(n')}}
      {\omega_{\rm pe}^2\eta_{l}^2+i\omega_{\rm RF}\nu-\omega_{\rm RF}^2}\ . 
\end{align}
The coefficients $\beta_{lm}^{(n)}$ and $\beta_{lm}^{(n')}$ are still not 
determined. They contain the information about the electrode configuration. 
Therefore, we have to distinguish between different probe\linebreak designs 
and focus on the IP and the MRP in the next two sections.

%%%%%%%%%%%%%%%%%%%%%%%%%%%%%%%%%%%%%%%%%%%%%%%%%%%%%%%%%%%%%%%%%%%%%%%%%%%%%%%%
%%%%%%%%%% Resonanzverhalten der Impedanzsonde %%%%%%%%%%
%%%%%%%%%%%%%%%%%%%%%%%%%%%%%%%%%%%%%%%%%%%%%%%%%%%%%%%%%%%%%%%%%%%%%%%%%%%%%%%%

\section{Admittance of the impedance probe}\label{sec:ResonanzIP}

In figure \ref{idealProbes} (left) the idealized impedance probe is depicted. 
It contains one spherical powered electrode $\set E_1$ and is, 
in our case, surrounded by a dielectric. The general current 
\eqref{strom3} is reduced to the current flowing to $\set E_1$, 
where the voltage $U_1$ is applied (For a shorter notation we substitute $\omega_{\rm RF}=\omega$ in the rest of the manuscript.)

\begin{equation}
i_1=Y_{11}U_{1}
   =i\omega\epsi_0\omega_{\rm pe}^2\sum_{l=0}^\infty\sum_{m=-l}^l
     \frac{l+1}{(R+\delta)^{2l+1}}
     \frac{\left(\beta_{lm}^{(1)}\right)^2 U_1}
          {\omega_{\rm pe}^2\eta_{l}^2+i\omega\nu-\omega^2}
\label{allgStromIP}
\end{equation}\\[-1cm]

\noindent The specified electrode configuration allows for $\beta_{lm}^{(1)}$ 
to be evaluated explicitly by its definition given in \eqref{FourierKoef}
\begin{equation}
\beta_{lm}^{(1)} 
=\frac{\sqrt{4\pi}}{\gamma_l}\,\delta_{l0}\delta_{m0}\ .
\label{BetaIP}
\end{equation}\\[-1cm]

\noindent Entering \eqref{BetaIP} into the current \eqref{allgStromIP} and 
exploiting the Kronecker deltas yields the admittance of the spherical 
impedance probe

\begin{equation}
Y_{\rm IP}
=\frac{4\pi\epsi_0\omega_{\rm pe}^2}{(R+\delta)\gamma_0^2}
 \frac{i\omega}
      {\omega_{\rm pe}^2\eta_{0}^2+i\omega\nu-\omega^2}
=\left(\frac{1}{i\omega C_0}
+\frac{i\omega+\nu}{\eta_{0}^2 \omega_{\rm pe}^2 C_0}\right)^{-1}\ .
\label{AdmittanzIP}
\end{equation}\\[-1cm]

\noindent It describes the coupling between the electrode and ground, which 
is in infinite distance to the probe. Equation \eqref{AdmittanzIP} 
shows that the impedance probe provides just one resonance mode 
with the resonance frequency $\omega_{00}$, which is given by the 
eigenvalue of the conservative operator $\op T_C$

\begin{equation}
\omega_{00} 
=\pm\omega_{\rm pe}
 \sqrt{1-\frac{\epsi_D\left(1-\frac{d}{R}\right)}
              {1+(\epsi_D-1)\left(1-\frac{d}{R}\right)}
 \left(1+\frac{\delta}{R}\right)^{-1}}\ .
\label{resIP}
\end{equation}\\[-1cm]

\noindent Neglecting the dielectric ($d=0$ and $\epsi_D=1$), the resonance 
frequency \eqref{resIP} reduces to the well known sheath resonance 
in spherical geometry \cite{harp1964b}. It can also be called 
``monopole resonance'' referring to the one electrode system. 

Different spectra of the impedance probe will be depicted and discussed within section \ref{sec:Comp}. We will compare them to the spectra of the multipole resonance probe and discuss advantages and disadvantages. The admittance of the MRP is determined in the next section.

%\pagebreak
%%%%%%%%%%%%%%%%%%%%%%%%%%%%%%%%%%%%%%%%%%%%%%%%%%%%%%%%%%%%%%%%%%%%%%%%%%%%%%%%
%%%%%%%%%% Admittance of the multipole resonance probe %%%%%%%%%%
%%%%%%%%%%%%%%%%%%%%%%%%%%%%%%%%%%%%%%%%%%%%%%%%%%%%%%%%%%%%%%%%%%%%%%%%%%%%%%%%

\section{Admittance of the multipole resonance probe}\label{sec:ResonanzMRP}

The idealized multipole resonance probe is shown in figure 
\ref{idealProbes} (right). The probe consists of an upper electrode 
$\set E_1$ and a lower electrode $\set E_2$, where the 
voltages $U_1$ and $U_2$ are applied, respectively. We calculate 
the current flowing to the electrode $\set E_1$
\begin{equation}
i_1
=\sum_{n'=1}^2 Y_{1n'}U_{n'}
=\sum_{n'=1}^2 \sum_{l=0}^\infty\sum_{m=-l}^l
 \frac{l+1}{(R+\delta)^{2l+1}}
 \frac{i\omega\epsi_0\omega_{\rm pe}^2\beta_{lm}^{(1)}\beta_{lm}^{(n')} U_{n'}}
      {\omega_{\rm pe}^2\eta_{l}^2+i\omega\nu-\omega^2}\ .
\label{allgStromMRP}
\end{equation}
Owing to the two different electrodes, we have to determine 
two coefficients $\beta_{lm}^{(1)}$ and $\beta_{lm}^{(2)}$
\begin{eqnarray}
\beta_{lm}^{(1)}
& = & \frac{\sqrt{(2l+1)\pi}}{\gamma_l}
       \frac{\sqrt{\pi}}{2\Gamma\left(1-\frac{l}{2}\right)\Gamma\left(\frac{3+l}{2}\right)}\,\delta_{m0}\ ,\\
\beta_{lm}^{(2)} 
& = & \frac{\sqrt{(2l+1)\pi}}{\gamma_l}
      \left(\frac{2\sin(l\pi)}{l\pi(1+l)}
     -\frac{\sqrt{\pi}}{2\Gamma\left(1-\frac{l}{2}\right)\Gamma\left(\frac{3+l}{2}\right)}\right)\,\delta_{m0}\ .
\end{eqnarray}
The gamma function $\Gamma\left(1-\frac{l}{2}\right)$, which is 
present in both coefficients, becomes infinity for all even 
$l>0$. Furthermore, the sine vanishes for all $l>0$. Thus, 
the coefficients vanish also for all even $l>0$ and the 
remaining coefficients for odd $l=2l'-1$ can be combined
\begin{equation}
\beta_{lm}^{(1/2)}
=\pm\frac{\sqrt{4l'-1}}{\gamma_{2l'-1}}
    \frac{\pi\,\delta_{l\, 2l'-1}\delta_{m0}}
         {2\Gamma\left(\frac{3}{2}-l'\right)\Gamma(l'+1)}
=\pm\beta_{2l'-1}\,\delta_{l\,2l'-1}\delta_{m0}\qkq l'\in\N \label{BetaMRP}\ .
\end{equation}
The positive sign belongs to $\beta_{lm}^{(1)}$ and the 
negative to $\beta_{lm}^{(2)}$. \eqref{BetaMRP} shows the influence 
of the symmetric geometry to the resonance modes of the probe-plasma 
system. The even modes, instead of $l=0$, vanish in the calculation. 
In the limit $l\rightarrow0$ the coefficients become equal
\begin{equation}
\beta_{00}^{(1)}=\frac{\sqrt{\pi}}{\gamma_0}=\beta_{00}^{(2)}\label{BetaMRP0}\ .
\end{equation}
Introducing \eqref{BetaMRP} and \eqref{BetaMRP0} into the 
current \eqref{allgStromMRP}, the current can be simplified %exploiting the Kronecker deltas
\begin{equation}
i_1
=\sum_{l'=1}^\infty
 \frac{2l'}{(R+\delta)^{4l'-1}}
 \frac{i\omega\epsi_0\omega_{\rm pe}^2\left(\beta_{2l'-1}\right)^2 
 \left(U_{1}-U_{2}\right)}
      {\omega_{\rm pe}^2\eta_{2l'-1}^2+i\omega\nu-\omega^2}  
+\frac{1}{R+\delta}
 \frac{i\omega\epsi_0\omega_{\rm pe}^2\left(\beta_{00}^{(1)}\right)^{2}
      \left(U_{1}+U_{2}\right)}
      {\omega_{\rm pe}^2\eta_{0}^2+i\omega\nu-\omega^2}\ .
\label{StromMRP}
\end{equation}
The MRP is operated with symmetric voltages $U=U_1=-U_2$. Owing to the electric symmetry the contribution of the zero mode vanishes and we evaluate the admittance %of the MRP
\begin{equation}
Y_{\rm MRP}
=\sum_{l'=1}^\infty
 \frac{4l'}{(R+\delta)^{4l'-1}}
 \frac{i\omega\epsi_0\omega_{\rm pe}^2\left(\beta_{2l'-1}\right)^2}
      {\omega_{\rm pe}^2\eta_{2l'-1}^2+i\omega\nu-\omega^2}
=\sum_{l'=1}^\infty 
 \left(\frac{1}{i\omega C_{l'}}
+\frac{i\omega+\nu}{\omega_{\rm pe}^2\eta_{2l'-1}^2 C_{l'}}\right)^{-1}\ .
\label{AdmittanzMRP}
\end{equation}
The second expression is identical to the admittance directly derived 
for the MRP \cite{lapke2008}. It shows that only odd modes of 
the probe-plasma system can be excited, where again the resonance 
frequencies are determined by the eigenvalues 
$\omega_{2l'-1}=\omega_{\rm pe}\eta_{2l'-1}$ of $\op T_C$. Due to the geometric and electric symmetry the electrodes couple to each other and not to ground.

\section{Comparison between IP and MRP}\label{sec:Comp}

\begin{figure}[h!]
\includegraphics[width= 0.8\columnwidth]{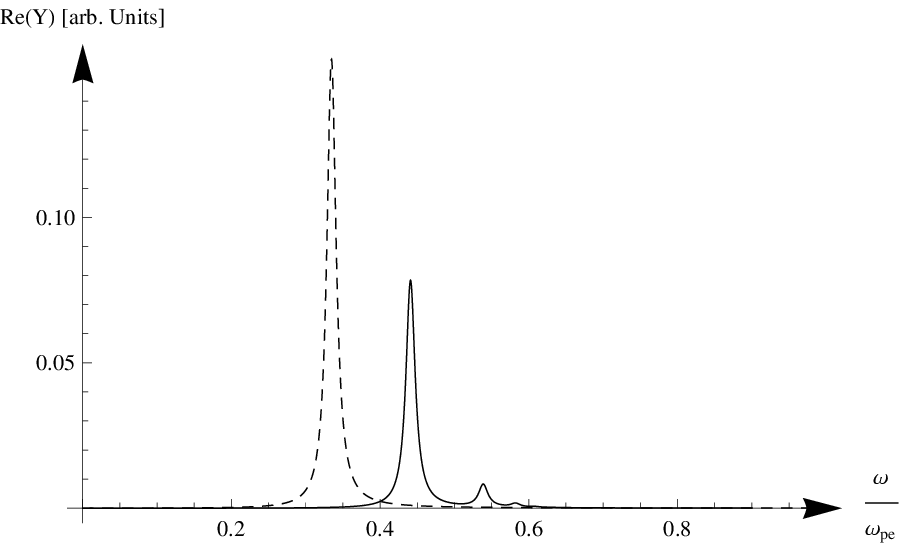}
\caption{Spectra of the impedance probe (dashed) and multipole resonance probe (solid) with the prototype parameter of the MRP $R=4\,{\rm mm}$, $d=1\,{\rm mm}$, $\varepsilon_D=4.6$ and assumed plasma parameter $\delta=0.2\,{\rm mm}$, $\omega_{\rm pe}=2\pi\,10^9\,{\rm s}^{-1}$, $\nu=0.015\omega_{\rm pe}$.}
\label{YProto}
\end{figure}

Within the last two sections the admittances of the impedance probe and multipole resonance probe are derived. Now, we compare their spectra based on the parameters of the MRP prototype: probe tip radius $R=4\,{\rm mm}$, thickness of the dielectric $d=1\,{\rm mm}$, and permittivity $\varepsilon_D=4.6$ \cite{lapke2011}. Additionally, we assume a sheath thickness of $\delta=0.2\,{\rm mm}$, a plasma frequency of $\omega_{\rm pe}=2\pi\,10^9\,{\rm s}^{-1}$, and a collision frequency of $\nu=0.015\omega_{\rm pe}$. Figure \ref{YProto} shows the corresponding spectra of both probes (IP dashed and MRP solid). Obviously, the resonance frequency of the IP is smaller than the ones of the MRP, as expected from the derived eigenvalues. Furthermore, one can see that the mode of the IP is less damped than the ones of the MRP. 

\begin{figure}[h!]
\includegraphics[width= 0.8\columnwidth]{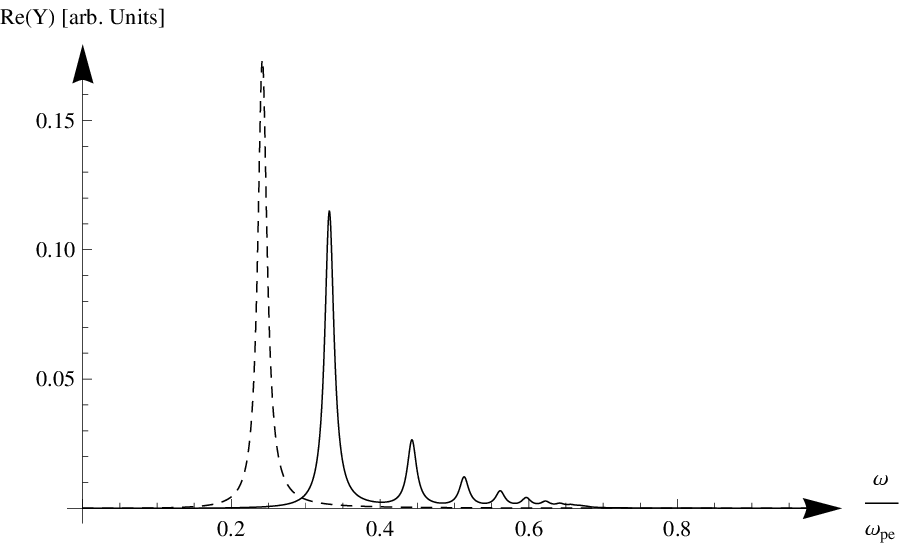}
\caption{Spectra of the impedance probe (dashed) and multipole resonance probe (solid) with the probe parameter $R=4\,{\rm mm}$, $d=0.2\,{\rm mm}$, $\varepsilon_D=4.6$ and assumed plasma parameter $\delta=0.2\,{\rm mm}$, $\omega_{\rm pe}=2\pi\,10^9\,{\rm s}^{-1}$, $\nu=0.015\omega_{\rm pe}$ and .}
\label{YDiel1}
\end{figure}

\begin{figure}[h!]
\includegraphics[width= 0.8\columnwidth]{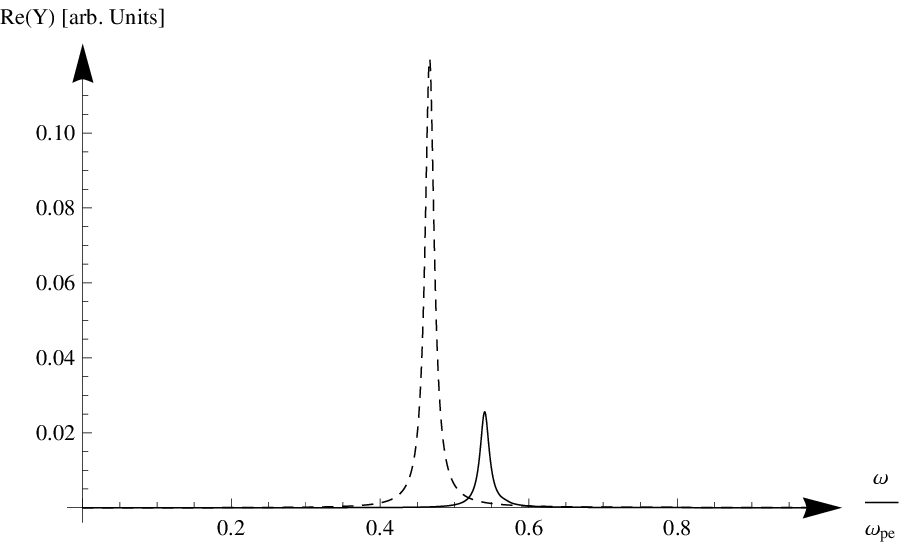}
\caption{Spectra of the impedance probe (dashed) and multipole resonance probe (solid) with the probe parameter $R=4\,{\rm mm}$, $d=2\,{\rm mm}$, $\varepsilon_D=4.6$ and assumed plasma parameter $\delta=0.2\,{\rm mm}$, $\omega_{\rm pe}=2\pi\,10^9\,{\rm s}^{-1}$, $\nu=0.015\omega_{\rm pe}$.}
\label{YDiel2}
\end{figure}

The spectra depict in figure \ref{YDiel1} are computed with a thickness of the dielectric that is equal to the sheath thickness $d=\delta=0.2\,{\rm mm}$. This leads to a shift of the resonance frequencies to smaller values and to less damping of the higher modes. Due to that, the contribution of the higher modes to the spectrum is observable. Increasing the thickness of the dielectric to $d=2\,{\rm mm}$ yields the spectra in figure \ref{YDiel2}, where the resonance frequencies are shifted to higher frequencies. The higher modes of the multipole resonance probe are not observable anymore. 

Both probes provide spectra with a dominant resonance peak. The monopole mode of the impedance probe is unique and just slightly damped by an increasing thickness of the dielectric. This allows a measurement also in a plasma with higher collision frequencies. However, the fact that the current couples to ground at infinity means that the interpretation of a measurement based on the excitation of the resonance mode has to be understood as an average reaction of the whole plasma. Therefore, the measurement is not local, which can be seen as a disadvantage of the impedance probe. Another advantage is the simple design.

Choosing an optimized thickness of the dielectric for the multipole resonance probe, the higher modes are strongly damped. This yields an uniquely observable dipole mode, as done for the prototype. In principle the MRP acts like a dipole with a rapid decreasing electric field, which provides a local measurement but the design is more complex to ensure the geometrical and electrical symmetry.

\section{Summary and conclusion}

Based on the result that the admittance of an electrostatic probe 
in arbitrary geometry is given by the resolvent of the dynamical 
operator $\op T_C+\op T_D$, we derived the general admittance 
in spherical geometry. Therefore, we determined the matrix 
representation of the resolvent by the eigenvalues and 
eigenfunctions of the conservative operator $\op T_C$. 
In general, the operators $\op T_C$ and $\op T_D$ do not commute. 
However, in spherical geometry they can be projected on the same subdomain with spherical 
harmonics as basisfunctions. This allows for the exact analytical 
representation of the general admittance in spherical geometry.

Two different probe designs were chosen to compare the general 
admittance with\linebreak established results: The spherical impedance 
probe and the multipole resonance probe. In both cases we showed 
that the admittances are simplified expressions of the general 
admittance. The corresponding resonance frequencies are given 
by the eigenvalues of the conservative operator $\op T_C$, 
respectively. Concerning the impedance probe, the resonance\linebreak 
frequency is equal to the sheath resonance and might also be 
called monopole resonance. The admittance of the multipole 
resonance probe is identical to the admittance, which is derived 
directly for the specific design \cite{lapke2008}.

Both probe designs provide spectra with a dominant resonance peak which is clearly detectable in a measurement. The impedance probe has a simple design and due to that always a unique resonance, but the measurement is not local. The multipole resonance probe has a more complex design to ensure the geometrical and electrical symmetry. Due to that symmetry the MRP acts like a dipole with a rapidly decreasing field, that the measurement is local.

The analytic solution presented here is restricted to spherical 
geometry. However, the solution strategy can also be performed in an arbitrary geometry. Therefore, an appropriate  set of orthonormal basis functions has to be found to determine an approximated matrix representation of the operators. This will lead to an efficient calculation of the approximated spectral response instead of a simulation. Possibly, a perturbation approach is useful to determine the admittance by the matrix representation of the resolvent.

\pagebreak

\acknowledgments
The authors acknowledge the support by the Federal Ministry of Education and Research (BMBF) in frame of the project PluTO, and also the support by the Deutsche Forschungsgemeinschaft (DFG) via Graduiertenkolleg GK 1051, Collaborative Research Center TRR 87, and the Ruhr University Research School. Gratitude is expressed
to M Lapke, C Schulz, R Storch, T Styrnoll, P Awakowicz, T Musch, and I Rolfes, who are part of the MRP-Team at the Ruhr University Bochum.

\pagebreak

\pagebreak

\pagebreak

\pagebreak

\pagebreak


\clearpage
\begin{thebibliography}{10}

\bibitem{takayama1960}
K. Takayama, H. Ikegami, und S. Miyazaki, Phys. Rev. Let. {\bf 5}, 238 (1960).


\bibitem{messiaen1966}
A. M. Messiaen and P. E. Vandenplas, J. Appl. Phys. {\bf 37}, 1718 (1966)


\bibitem{waletzko1967}
J. A. Waletzko and G. Bekefi, Radio Sci. {\bf 2}, 489 (1967). 


\bibitem{vernet1975}
N. Vernet, R. Manning, and J. L. Steinberg, Radio Sci. {\bf 10}, 517 (1975). 


\bibitem{blackwell2005}
D. D. Blackwell, D. N. Walker, and W. E. Amatucci, Rev. Sci. Instrum {\bf 76}, 023503 (2005).


\bibitem{sugai1999} H. Kokura, K. Nakamura, I.P. Ghanashev, and H. Sugai, Japan. J. Appl. Phys {\bf 38}, 5262 (1999).


\bibitem{scharwitz2009} 
C. Scharwitz, M. B\"oke, J. Winter, M. Lapke, T. Mussenbrock, and R. P. Brinkmann, Appl. Phys. Lett. {\bf 94}, 011502 (2009). 


\bibitem{lapke2008} 
M. Lapke, T. Mussenbrock, and R. P. Brinkmann, Appl. Phys. Lett. { \bf 93}, 051502 (2008).


\bibitem{fejer1964}
J. A. Fejer, Radio Sci. {\bf 68D}, 1171 (1964)


\bibitem{harp1964}
R. S. Harp, Appl. Phys. Lett. {\bf 4}, 186 (1964)


\bibitem{crawford1964}
R. S. Harp and F. W. Crawford, J. Appl. Phys. {\bf 35}, 3436 (1964)


\bibitem{dote1965}
T. Dote and T. Ichimiya, J. Appl. Phys. {\bf 36}, 1866 (1965).


\bibitem{kostelnicek1968}
R. J. Kostelnicek, Radio Sci. {\bf 3}, 319 (1968).


\bibitem{cohen1971}
 A. J. Cohen and G. Bekefi, Phys. Fluids {\bf 14}, 1512 (1971).
 
 
\bibitem{tarstrup1972}
J. Tarstrup and W. J. Heikkila, Radio Sci. {\bf 4}, 493 (1972).


\bibitem{aso1973}
T. Aso, Radio Sci. {\bf 8}, 139 (1973).


\bibitem{bantin1974}
C. C. Bantin and K. G. Balmain, Can. J. Phys. {\bf 52}, 291 (1974).


\bibitem{booth2005} S. Dine, J.P. Booth, G.A Curley, C.S. Corr, J. Jolly, and J. Guillon, Plasma Scources Sci. Technol. {\bf 14}, 777 (2005).


\bibitem{walker2006}
D. N. Walker, R. F. Fernsler, D. D. Blackwell, W. E. Amatucci, and S. J. Messer, Phys. Plasmas {\bf 13}, 032108 (2006).


\bibitem{lapke2007}
M. Lapke, T. Mussenbrock, R. P. Brinkmann, C. Scharwitz, M. B\"oke, and J. Winter, Appl. Phys. Lett. {\bf 90}, 121502 (2007).


\bibitem{xu2009} J. Xu, K. Nakamura, Q. Zhang, and H. Sugai, Plasma Sources Sci. Technol. {\bf 18} 045009, (2009). 


\bibitem{xu2010} J. Xu, J. Shi, J. Zhang, Q. Zhang, K. Nakamura, and H. Sugai, Chinese Phys. B {\bf 19} 075206, (2010). 


\bibitem{li2010} B. Li, H. Li, Z. Chen, J. Xie G. Feng, and W. Liu, Plasma Sci. Technol. {\bf 12}, 513 (2010).


\bibitem{liang2011} I. Linag, K. Nakamura, and H. Sugai, Appl. Phys. Express {\bf 4}, 066101 (2011)


\bibitem{lapke2013}
M. Lapke, J. Oberrath, T. Mussenbrock und R. P. Brinkmann, Plasma Scources Sci. Technol. {\bf 22}, 025005 (2013).


\bibitem{harp1964b}
R. S. Harp, Appl. Phys. Lett.  {\bf 4}, 186 (1964).

\bibitem{lapke2011}
M. Lapke et al., Plasma Sources Sci. Technol. {\bf 20} 042001, (2011). 
\end{thebibliography}
\end{document}